\documentclass[a4paper,11pt]{article}
\pdfoutput=1 

\usepackage{jcappub} 

\usepackage[T1]{fontenc} 

\usepackage{multirow} 
\usepackage{times}

\title{\boldmath 
Angular Baryon Acoustic Oscillation measure at z=2.225 from the SDSS quasar survey
}

\author[a,b]{E. de Carvalho,}
\author[a]{A. Bernui,}
\author[c]{G. C. Carvalho,}
\author[a]{C. P. Novaes,}
\author[c]{H. S. Xavier}

\affiliation[a]{Observat\'orio Nacional, 20921-400, Rio de Janeiro, RJ, Brazil}
\affiliation[b]{Universidade\ do\ Estado\ do\ Amazonas,\ 69640-000,\ Tabatinga,\ AM,\ Brazil}
\affiliation[c]{Instituto de Astronomia, Geof\'isica e Ci\^encias Atmosf\'ericas, Universidade de 
S\~ao Paulo, Rua do Mat\~ao, 1226, 05508-090, S\~ao Paulo, SP, Brazil}

\emailAdd{edilsonfilho@on.br}
\emailAdd{bernui@on.br}
\emailAdd{gabriela@on.br}
\emailAdd{camilanovaes@on.br}
\emailAdd{hsxavier@if.usp.br}

\abstract{
Following a quasi model-independent approach we measure the transversal BAO mode 
at high redshift using the two-point angular correlation function (2PACF). 
The analyses done here are only possible now with the quasar catalogue from the twelfth data release 
(DR12Q) from the Sloan Digital Sky Survey, because it is spatially dense enough to allow the 
measurement of the angular BAO signature with moderate statistical significance and acceptable 
precision. 
Our analyses with quasars in the redshift interval $z \in [2.20,2.25]$ produce the angular BAO scale 
$\theta_{\mbox{\footnotesize\sc bao}} = 1.77^{\circ} \pm 0.31^{\circ}$ with a statistical significance of 
$2.12 \,\sigma$ (i.e., 97\% confidence level), calculated through a likelihood analysis performed using 
the theoretical covariance matrix sourced by the analytical power spectra expected in the 
$\Lambda$CDM concordance model. 
Additionally, we show that the BAO signal is robust --although with less statistical significance-- 
under diverse bin-size choices and under small displacements of the quasars' angular coordinates. 
Finally, we also performed cosmological parameter analyses comparing the 
$\theta_{\mbox{\footnotesize\sc bao}}$ predictions for wCDM and w$(a)$CDM models with 
angular BAO data available in the literature, including the measurement obtained here, jointly with 
CMB data. 
The constraints on the parameters $\Omega_M$, w$_0$ and w$_a$ are in excellent agreement with 
the $\Lambda$CDM concordance model. 
}

\begin{document}
\maketitle
\flushbottom

\section{Introduction} \label{sec1}
\paragraph{}
One of the most iconic and incomprehensible components of the universe is the {\em dark energy}, 
an unknown entity representing $\sim 70\%$ of the cosmos~\cite{PLA-XIII}. 
A powerful approach to unveil the dark energy mysteries is searching for the Baryon Acoustic 
Oscillations (BAO) signature at different epochs to probe its time 
evolution~\cite{PeeblesYu,SZ,seo-eisenstein,weinberg,eisenstein98,Meiksin}. 

The BAO is a primordial phenomenon that imprinted a characteristic scale, the sound horizon $r_{s}$, 
in the spatial distribution of cosmic objects. 
Almost a decade of success mapping the large-scale distribution of luminous matter provided large, 
dense, and deep surveys of cosmological tracers of the universe evolution like luminous red galaxies, 
emission line galaxies, and quasars~\cite{SDSS,WIGGLEZ,6dFGRS}. 
With these data, analyses of the two-point correlation function (2PCF) produced robust detections of 
the BAO signature in different data ensembles giving precise 
measurements of the distance-redshift relationship, mainly using luminous red 
galaxies~\cite{Eisenstein05,Cole05,Wang}, but also with clusters of galaxies~\cite{Hong,Veropalumbo}.

Moreover, to probe the early time evolution of the dark energy it is imperative to investigate the highest 
redshift available cosmological tracer, that is, the quasars~\cite{McDonald,White,Bassett}. 
The first BAO analyses with quasars, at $2.1 \le z \le 3.5$, used the data release 9 
(DR9) of the Baryon Oscillation Spectroscopic Survey, part of the Sloan Digital Sky Survey 
(SDSS)~\cite{SDSS}, to measure the BAO features in the three-dimensional correlation function in 
the Lyman-$\alpha$ flux fluctuations resulting in constraints on the angular diameter distance, $D_A$, 
and on the Hubble expansion rate, $H$, at the redshifts $z = 2.3$~\cite{Busca} and 
$z = 2.4$~\cite{Slosar}. 
More recent studies of the DR11~\cite{Delubac} and DR12~\cite{Bautista} quasar catalogues 
achieved more precise measurements of these quantities at $z = 2.34$ and $z=2.33$, respectively 
(see also~\cite{Ata}).

A common feature in these analyses is the assumption of a fiducial cosmological model to compute 
the comoving distance between pairs of astrophysical sources, and then calculate the 2PCF in order 
to analyze the excess probability of finding pairs at a given separation distance. 
In this case, the sample contains both the transversal and the radial BAO modes, and the traditional 
way to quantify the cosmic evolution is through the spherically-averaged distance, $D_V$, 
with mixed information from both modes~\cite{Eisenstein05}.

Alternatively,  we follow a {\em quasi} model-independent approach to measure the transversal BAO 
mode at high redshift, and for this we consider a sample of quasars located in a thin range of 
redshift of the SDSS DR12 dataset~\cite{Paris}. 
The main idea behind this approach is that, if the redshift bin is sufficiently thin then the two-point 
angular correlation function (2PACF), calculated over the angular coordinates without using a fiducial 
cosmological model, successfully captures the transversal BAO mode. 
Rigorously, this procedure is {\em quasi} model-independent because a cosmological model is used 
just in two steps~\cite{Sanchez11,Carnero,Carvalho,Salazar,Alcaniz}: 
(i) the transversal BAO analysis is done in a thin but finite redshift bin, i.e., $\delta z \ne 0$, and 
    therefore there is a small shift in the BAO angular position; however, for the quasar sample studied 
    here, in which $\delta z = 0.05$, the shift correction is only $0.9\%$ of the BAO measured scale 
    (as we shall show); 
(ii) the comoving sound horizon, $r_{s}$, used to calculate the angular diameter distance from the BAO 
     angular position, $D_A$, is obtained assuming a cosmological model (see, 
     e.g.,~\cite{Heavens,Alcaniz} for a discussion about this point). 

This work is organized as follows. 
In section~\ref{sec2} we briefly describe the quasar sample investigated, while in section~\ref{sec3} 
we explain the methodology of our analyses. 
In section~\ref{sec4} we show our results and discussions.
The conclusions and final remarks are presented in section~\ref{sec5}. 
We leave complementary analyses and robustness tests to the Appendix section.

\section{The Quasars data set} \label{sec2}
\paragraph{}
The data used in this work is part of the twelfth public Data Release Quasar catalogue (DR12Q) 
from phase III of the Sloan Digital Sky Survey (SDSS-III)~\cite{Eisenstein2011}
~\footnote{\url{www.sdss.org/dr12/algorithms/boss-dr12-quasar-catalog/}}. 
The complete sample of DR12Q contains $297,301$ quasars from the Baryon Oscillation Spectroscopic 
Survey (BOSS)~\cite{Dawson2013}, among which $184,101$ have $z \ge 2.15$ (actually, more than 
90\% of them are new discoveries), covering a total area of $9,376 \, \mbox{deg}^{2}$ in the sky. 
The full sample has been spectroscopically confirmed based in a visual inspection of the spectra of 
each quasar. 
The SDSS-III/BOSS limiting magnitudes for quasar target selection are $r \le 21.85$ or 
$g \le 22$~\cite{Paris}. 
Perhaps, the main challenge faced in the quasar BOSS survey was to obtain a high number-density 
sample, satisfying the proposed minimum threshold of 15 quasars per square degree~\cite{Paris}. 
In fact, the current sample is actually dense enough to allow for the analyses in very thin redshift 
bins. 
The methodology followed to retrieve this sample was based in a SDSS pipeline and is fully 
described in P\^aris et al. (2017)~\cite{Paris}.

With this exceptional catalogue, DR12Q, we search for a statistically significant angular BAO 
detection. 
The DR12Q sample is distributed between two disconnected regions of the sky, 
so we selected that one containing the largest amount of quasars, that is, the quasars located in the 
sky patch with $90^{\circ} < \alpha < 270^{\circ}$ (i.e., the North Galactic region), where $\alpha$ 
is the right ascension in equatorial coordinates; moreover, we first considered the quasar sample in 
the redshift range $2.20 - 2.80$. 
After several analyses that considers the large number of quasars (to minimize the statistical noise) 
in the thinest redshift bin (to minimize the non-linear contributions due to the projection effect, see, 
e.g., ref.~\cite{Sanchez11}), 
we finally selected the thin shell: $z \in [2.20, 2.25]$, which contains a total of $10,526$ quasars.

\section{\label{sec3} The two-point correlation functions}

Traditionally, the BAO analyses assume a fiducial cosmology to compute the comoving distance 
between pairs, then the characteristic scale is found through the three-dimensional (3D) two-point 
correlation function. 
Here instead, we use the two-dimensional (2D) version of this correlation function, that is, the 
two-point angular correlation function (2PACF), using quasars located in a thin redshift shell. 
This analysis is only possible now because the current dataset is spatially dense enough to 
allow for the measurement of the angular BAO signal with high statistical significance.

\subsection{The two-point angular correlation function}

\noindent
One way to characterize the clustering of galaxies, quasars, and other cosmological tracers is through 
the two-point correlation function (2PCF)~\cite{Peebles-Hauser,Davis-Peebles,Hewett,Hamilton,%
Landy-Szalay}. 
The 2PCF is based on counting pairs of cosmic objects in a data set (quasars, for example) $DD(s)$ 
at a given comoving 3D distance $s$ and compare this with the number of pairs $RR(s)$ from a 
random set. 

The most used 2PCF estimator in astrophysical applications is the Landy-Szalay 
(LS)~\cite{Landy-Szalay}, because it returns the smallest deviations for a given cumulative probability, 
besides to have no bias and minimal variance~\cite{Kerscher}. 
This estimator is defined by 
	\begin{eqnarray}\label{2PCF}
	\xi(s) \equiv \frac{DD(s) - 2DR(s) + RR(s)}{RR(s)},
	\end{eqnarray} 
where $DR(s)$ counts the pairs, one in the data set and the other in the random set, separated by a 
distance $s$. 
The quantity $\xi(s)$ gives the excess probability of finding two points of a data set at a given 
separation distance $s$ when compared to a completely homogeneous distribution. 
If there is an excess of probability localized around a characteristic scale, as in the case of acoustic 
scale, it will appear as a bump in the $\xi(s)$ vs. $s$ plot. 
The location of the bump indicates the statistically preferred distance between the pairs of the studied 
sample. 

A 3D 2PCF estimator assumes a cosmological model (flat $\Lambda$CDM, for example) to calculate 
the comoving distance $s$ between pairs of the data and the random catalogues. 
To avoid this model dependence we follow a different approach. 
We use the two-point angular correlation function 2PACF~\cite{PeeblesYu} to estimate the transversal 
BAO contribution, analysis that is possible to be done selecting the data sample in a sufficiently thin 
redshift shell $\delta z$. 
The expression for the 2PACF estimator, $\omega(\theta)$, at mean redshift $\overline{z}$, is given by
\begin{eqnarray}\label{2PACF}
\omega(\theta) \equiv \frac{DD(\theta)-2DR(\theta)+RR(\theta)}{RR(\theta)} \, ,
\label{eq2}
\end{eqnarray}
with $\theta$, the angular separation between any pair of quasars, given by 
\begin{eqnarray}	
\,\,\theta\,=\,\arccos[\sin{\delta_{A}}\sin{\delta_{B}}+
\cos{\delta_{A}}\cos{\delta_{B}}\cos(\alpha_{A}-\alpha_{B})] \, , \nonumber
\end{eqnarray}
where $\alpha_{A}, \alpha_{B}$ and $\delta_{A}, \delta_{B}$ are the right ascension and declination 
coordinates of the quasars $A$ and $B$, respectively~\cite{Landy-Szalay,Sanchez11,Crocce11a}. 

To find the angular scale $\theta_{\mbox{\footnotesize\sc fit}}$ of the BAO bump in the 2PACF 
we use the method proposed in S\'anchez et al., 2011~\cite{Sanchez11}, which is based on the 
empirical parametrization of $\omega = \omega(\theta)$ 
\begin{eqnarray}\label{ajuste}			    
\omega(\theta) = A + B\theta^{\,\gamma} + C\exp^{-(\theta-\theta_{\rm FIT}) / 2\sigma_{\rm FIT}^{2}} \, ,
\label{eq3}
\end{eqnarray}
where $A$, $B$, $C$, $\gamma$, $\theta_{\mbox{\footnotesize\sc fit}}$, and 
$\sigma_{\mbox{\footnotesize\sc fit}}$ are free parameters. 
Therefore, the 2PACF best-fitting empirical expression (Eq.~\ref{eq3}) provides 
$\theta_{\mbox{\footnotesize\sc fit}}$, while the width of the bump, 
$\sigma_{\mbox{\footnotesize\sc fit}}$, 
define the error associated to the measured $\theta_{\mbox{\footnotesize\sc fit}}$. 
The final determination of the acoustic peak is achieved after accounting for an effect that produces 
a small displacement of the BAO bump, as we discuss below. 

The finite thickness $\delta z$ of the shell containing the data produces a shift in the acoustic peak 
due to a {\em projection effect}. 
To understand this effect consider first all quasars on a spherical shell with radius equal to the 
characteristic BAO scale and centred on another quasar, therefore  
contributing to the BAO bump in a 3D analysis, where the central quasar is located at the redshift 
$\overline{z}$ (the mean value of the data in the redshift bin, of width $\delta z$, in study). 
In the case of the 2D analysis, the BAO signature in the 2PACF comes from the quasars displayed 
along circles, or quasi circles, located in the transversal plane (the plane perpendicular to the 
line-of-sight). 
This means that, in the 2D analysis, one is assuming that all quasars in the redshift bin are projected 
onto the plane with redshift $\overline{z}$. 
The projection effect produced by this assumption has been studied in detail (see, 
e.g.,~\cite{Sanchez11,Carvalho}) and the net result is a shift in the angular position of the BAO 
bump, that can be estimated using numerical analysis. 
S\'anchez et al. (2011) have shown that for small $\delta z$ values and for data at $\overline{z} > 2$, 
which is our case, the dependence of the shift on the cosmological parameters is negligible. 
In the following section we perform such detailed numerical analysis that confirms this prediction 
for our quasars sample.

\subsection{The dependence of the angular BAO scale with the fiducial model} 

To compute the dependence of the angular BAO scale with the fiducial cosmology one has to 
perform a numerical integration to evaluate the projection effect on the expected 2PCF, $\xi_E$. 
For this, consider a sample of cosmic objects in a thin redshift bin $z \in [z_1, z_2]$, that is, 
$z_1 \simeq z_2$. 
Because of this, $\xi_E^{z_1}(s) \simeq \xi_E^{z_2}(s) \simeq  \xi_E^{\bar{z}}(s)$, 
where $\overline{z} \equiv (z_{1}+z_{2})/2$. 
Therefore, one can obtain the expected 2PACF, $\omega_E^{\bar{z}}(\theta)$, as a projection of the 
expected 2PCF, $\xi_E^{\bar{z}}(s)$, in that redshift shell 
\begin{eqnarray}			    
\omega_E^{\bar{z}}(\theta)=\int_{0}^{\infty}dz_{1}\, \phi(z_{1})
\int_{0}^{\infty}dz_{2}\, \phi(z_{2})\, \xi_E^{\bar{z}}(s) \, ,
\label{eq4}
\end{eqnarray}
where $\phi$ is the top-hat selection function which should be normalized to 1, 
$s$ is the 3D separation between pairs, and for a spatially flat Robertson-Walker metric 
is calculated using the relation: 
$s = \sqrt{\zeta^2(z_1)+\zeta^2(z_2)-2\zeta(z_1)\zeta(z_2) \cos{\theta}}$, 
where $\theta$ is the angular separation between those pairs, and $\zeta(z_i)$ is the comoving radial 
distance to the quasar with redshift $z_i$, obtained using a cosmological model. 
The expected 2PCF is given by 
\begin{eqnarray}			    
\xi_E^{\bar{z}}(s)=\int_{0}^{\infty}\frac{dk}{2\pi^{2}}\,k^{2}\,j_{0}(ks)\,P(k,\overline{z}) \, ,
\end{eqnarray} 
where $j_{0}$ and $P(k,\overline{z})$ are the spherical Bessel function of zero order and the matter power 
spectrum, respectively. 
To calculate the matter power spectrum one needs to assume a cosmological model, and 
this is the model dependence step mentioned in section~\ref{sec1}. 
However, as we shall confirm in section~\ref{IVc}, for the features of the sample in study, 
$\overline{z} = 2.225$ and $\delta z = 0.05$, the model dependence of our BAO measurement is 
actually weak~\cite{Sanchez11}.

\section{\label{sec4} Data analyses and Results} 

In this section we perform the analysis that lead us to a robust measurement of the angular BAO 
scale in the DR12Q sample.

\subsection{The 2PACF and the bin-size selection}\label{bin-size}

Note that the 2PACF is estimated for equally spaced values of $\theta$ in the range 
$0^{\circ} \leq \theta \leq {10^{\circ}}$, in a total of $N_{b}$ bins. 
However, to extract the BAO bump information we are looking for, $\theta_{\mbox{\footnotesize\sc fit}}$ 
and $\sigma_{\mbox{\footnotesize\sc fit}}$, we fit the equation~(\ref{ajuste}) only to the points in the 
range $0^{\circ} < \theta \leq {3.5^{\circ}}$, well beyond the angular BAO scale. 

\begin{figure*}
\mbox{\hspace{-0.6cm}
\includegraphics[width=5.8cm, height=4.7cm]{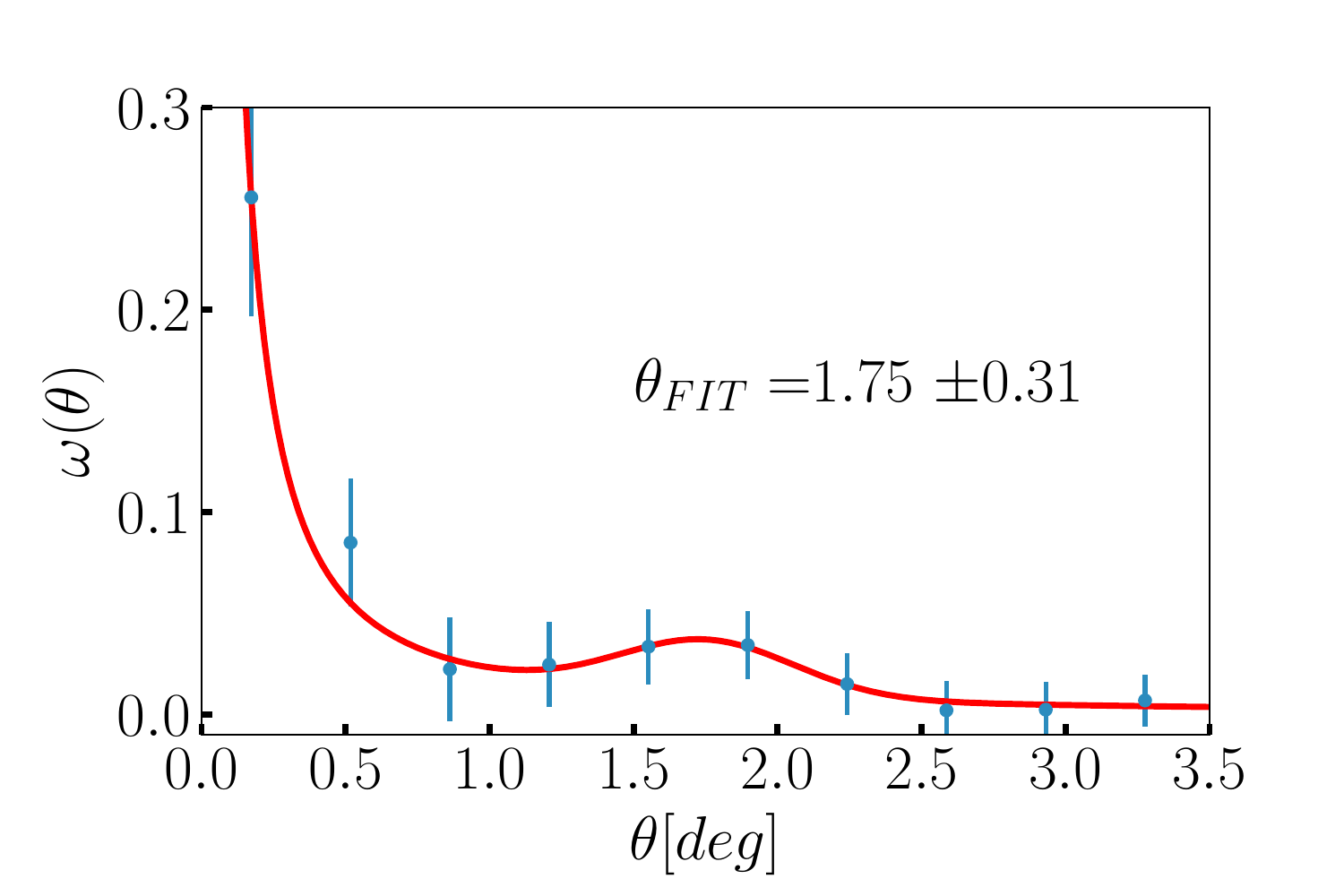}
\hspace{-0.7cm}
\includegraphics[width=5.8cm, height=4.7cm]{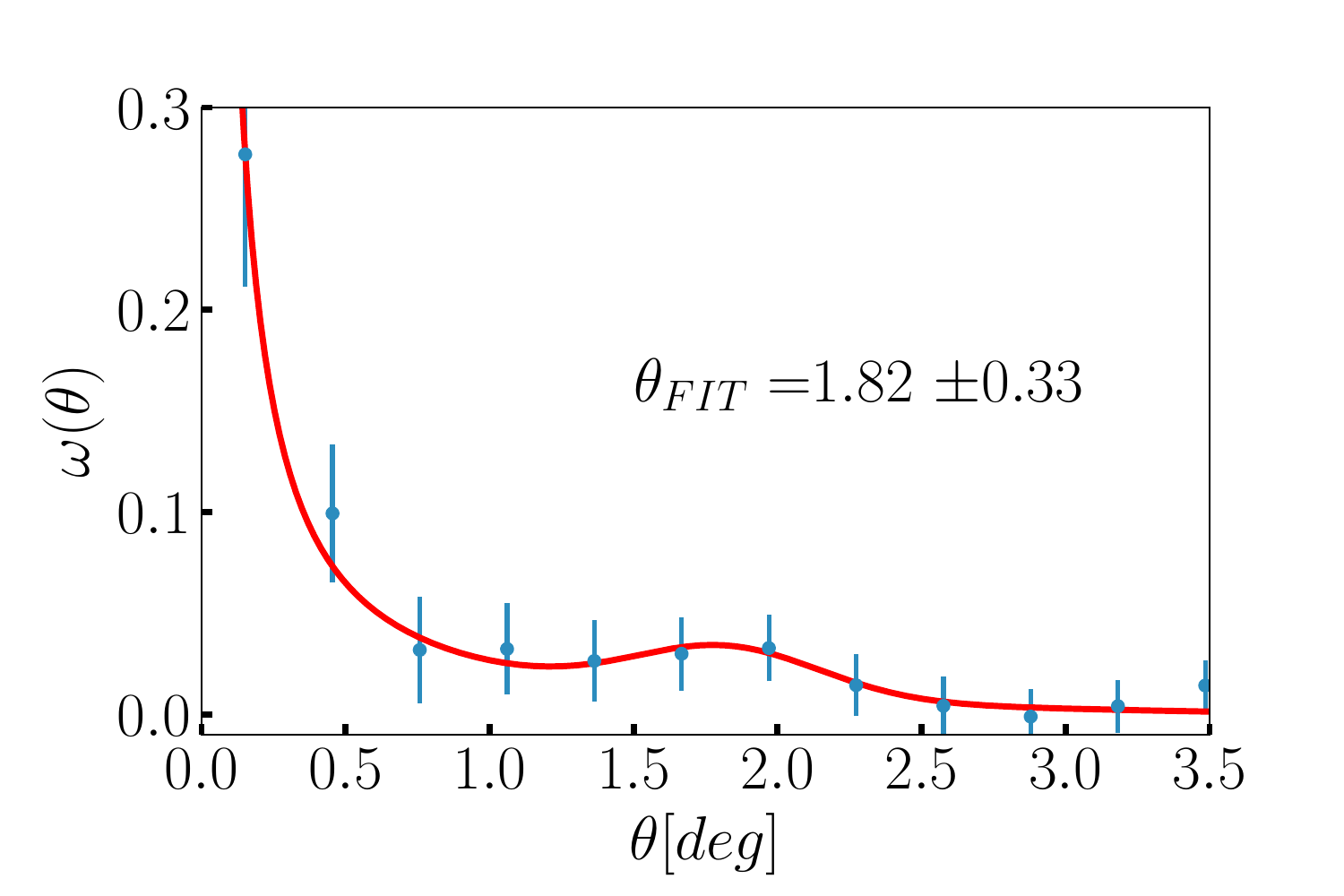}
\hspace{-0.7cm}
\includegraphics[width=5.8cm, height=4.7cm]{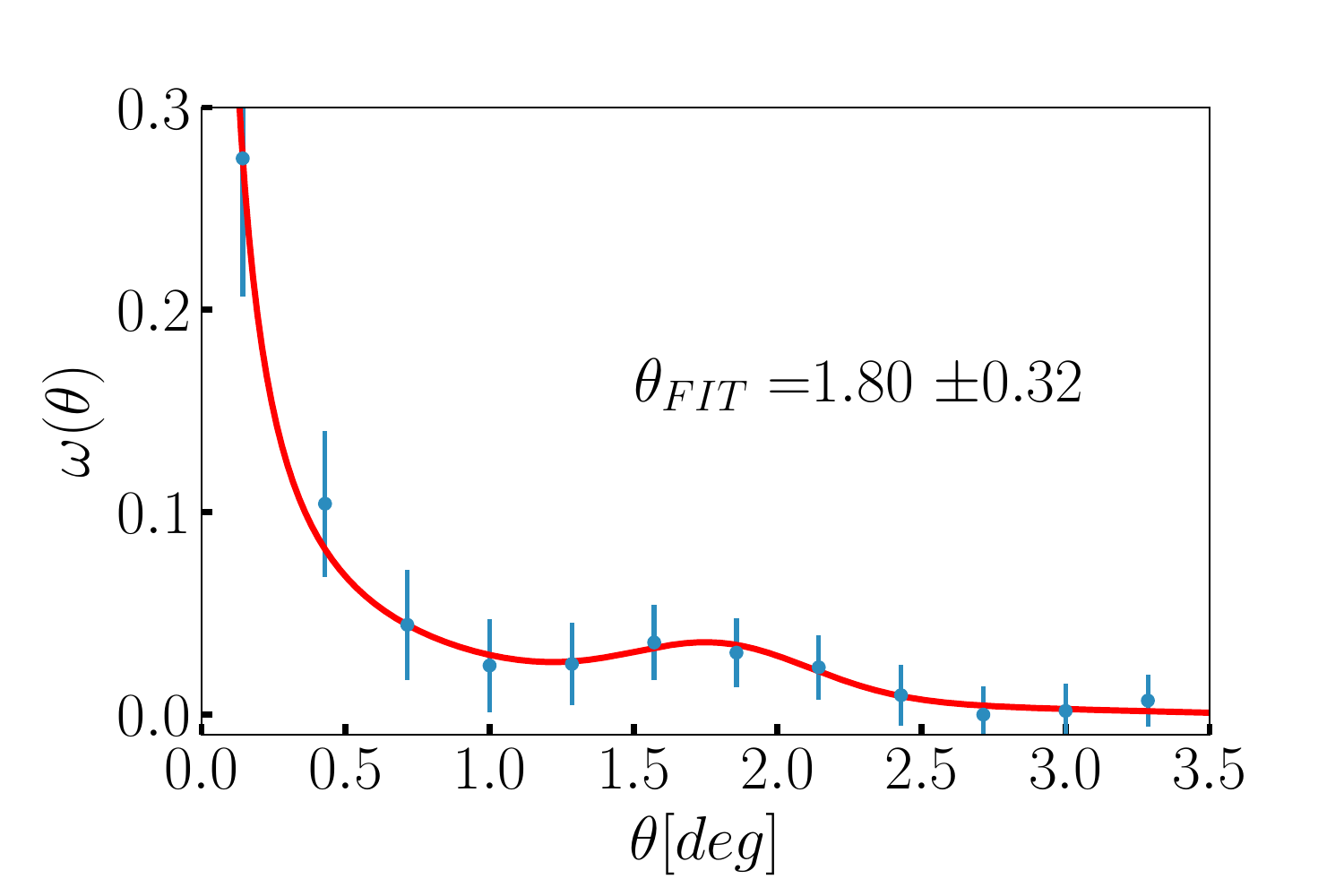}
}
\vspace{-0.5cm}
\caption{The 2PACF calculated from the sample DR12Q at $\overline{z} = 2.225$ (dots) and 
the best-fit curve obtained using the equation (\ref{ajuste}) (continuous line). 
In this case we have used $N_{b}=29$ (left panel), $N_{b}=33$ (center panel), and $N_{b}=35$ (right 
panel), see the text for details.} 
\label{fig1}
\end{figure*}

In order to verify the influence of the binning of the angular separation $\theta$ on our result, 
we tested $N_{b} = 29, \,33$, and 35 --as presented in figure~\ref{fig1}-- as well as several other 
numbers of bins. 
The analyses for $N_{b} =$ 29, 33, and 35, result in the values 
$\theta_{\mbox{\footnotesize\sc fit}}=1.75^{\circ}\pm0.31^{\circ}; 1.82^{\circ}\pm0.33^{\circ};$ and 
$1.80^{\circ}\pm0.32^{\circ}$, respectively. 
These analyses show that $\theta_{\mbox{\footnotesize\sc fit}}$ slightly depends on the choice of 
$N_{b}$. 
The criterion to pick the best choice for $N_{b}$ is through the statistical significance value, which is a way to discriminate the measurement with the best signal to noise ratio. 
In the following analyses we show that the largest statistical significance is obtained with 
$N_{b}=29$, as summarized in table~\ref{table1}.

\subsection{The 2PACF: the error bars calculation and the statistical significance}\label{errors}

Given the DR12Q sample, in the redshift range $z = 2.20 - 2.25$ and mean redshift 
$\overline{z}=2.225$, we use the LS estimator given by equation (\ref{eq2}), to obtain the 2PACF, 
$\omega(\theta)$, as displayed in figure~\ref{fig1}. 
The calculation of the data points  $\omega(\theta)$, represented as dots in the plots of 
figure~\ref{fig1}, and their error bars are as follows.

Consider that we have $N$ random catalogues, each one sharing the same features as the data sample, 
that is, it contains the same number of objects and geometry of the data survey. 
Assume also that we perform this calculation considering $N_b$ bins. 
Using equation~(\ref{eq2}), we calculate $N$ times the 2PACF, $\{ \omega^j(\theta) \},$ with $j=1, \cdots, N$, 
one for each random catalogue. 
Then, the 2PACF is achieved, for each bin $k$, with $k=1, \cdots, N_b$, from the arithmetic mean: 
$\omega(\theta_{k}) = (1/N) \, \sum_{j=1}^N \, \omega^j(\theta_{k})$. 

The calculation of the error bars of the 2PACF data points is a sensitive issue. 
The correct calculation of these errors is either through simulated mocks or equivalently, through the 
theoretical approach (see, e.g.,~\cite{Crocce11a,Crocce11b,Norberg,Friedrich}). 
In fact, here we obtain them following the analytical covariance matrix method as described in 
ref.~\cite{Crocce11b}. 

Firstly, we obtain the analytical angular power spectrum $C_{\ell}$ using the {\sc CAMBsources} 
code~\cite{Challinor}\footnote{\url{http://camb.info/sources}}, adopting the following setup: \\
--\, a maximum $\ell$ of 720, enough to sample the 2PACF with our binning choice; \\
--\, the $\Lambda$CDM fiducial cosmology, according to Planck's second data 
release~\cite{PLA-XIII}; \\
--\, a top-hat redshift window of width $\delta z = 0.05$, centred at $z=2.225$, representing the 
narrow spectroscopic redshift bin analysed here; \\
--\, a linear bias of 4.25, compatible with the one measured for quasars; \\
--\, we included redshift space distortions and gravitational lensing effects, and computed both the  
linear and non-linear $C_{\ell}$.

\begin{figure*}[h!]
\hspace{-0.95cm}
\includegraphics[width=0.61\textwidth]{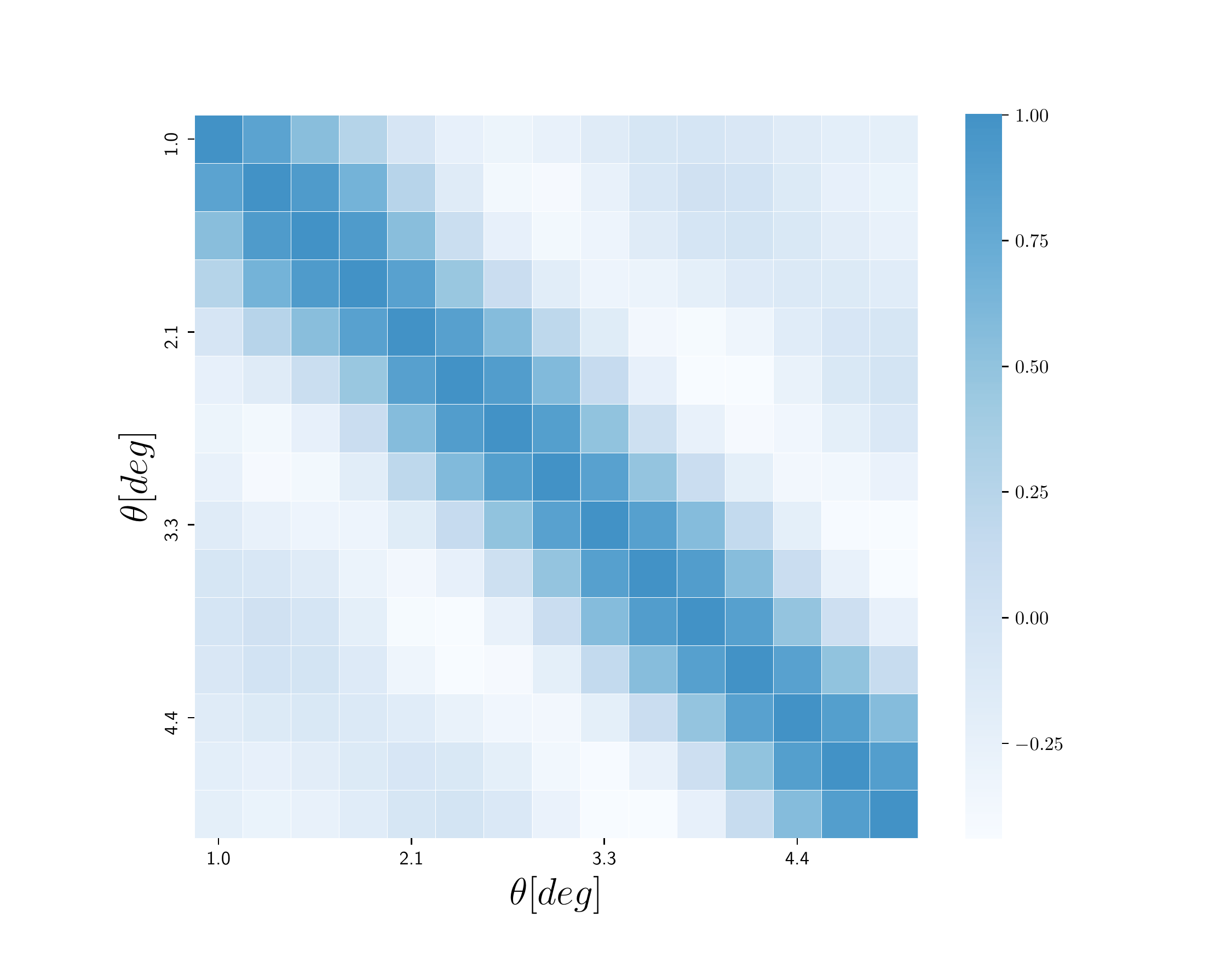}
\hspace{-1.5cm}
\includegraphics[width=0.61\textwidth]{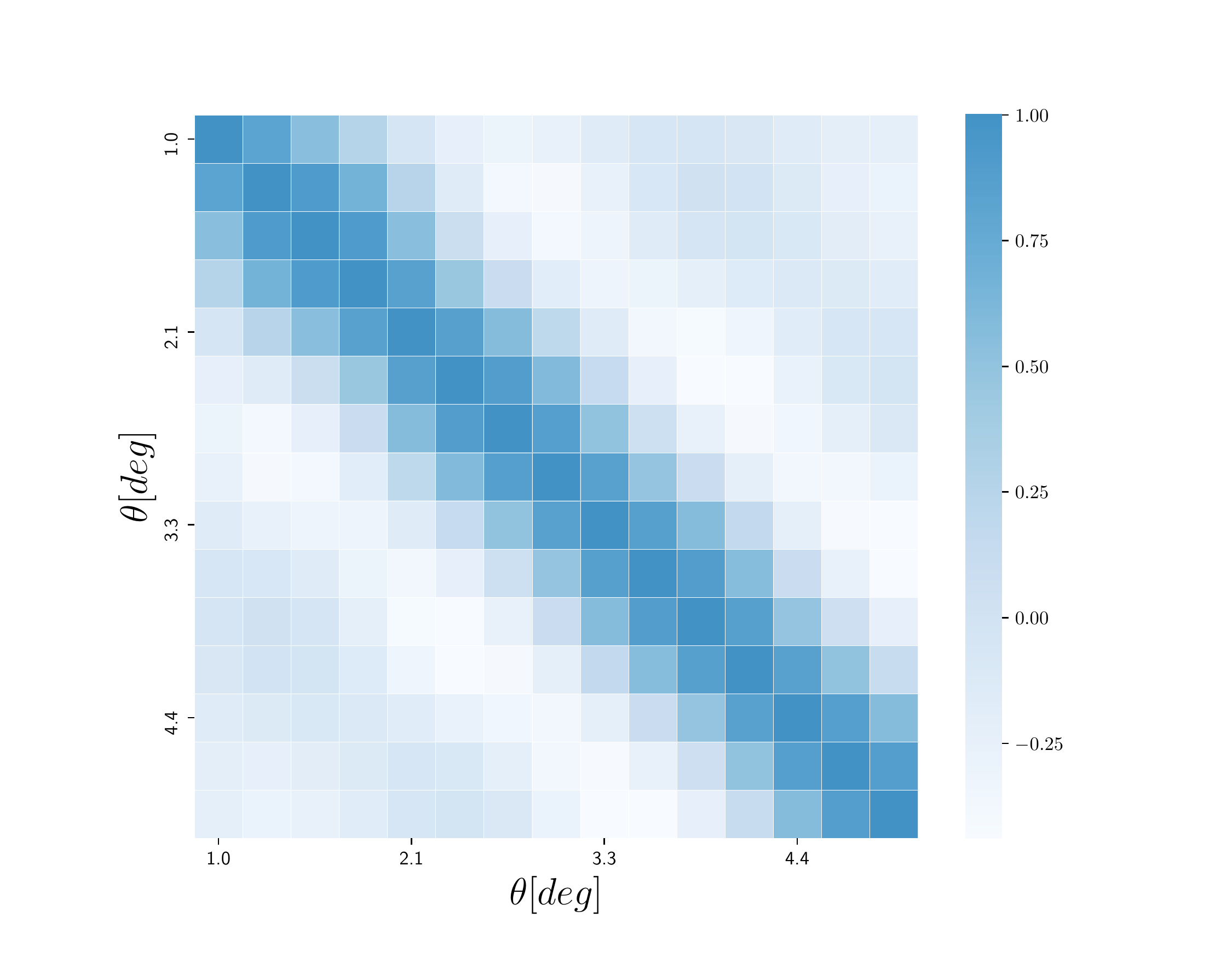}
%
\caption{
Analytical covariance matrices, given by eq.~\ref{eqCov}, in the linear (left panel) and non-linear 
(right panel) cases.
}
\label{fig2}
\end{figure*}

\noindent
Then we obtain the analytical covariance matrix expected in the fiducial cosmology $\Lambda$CDM 
\begin{eqnarray}\label{eqCov}
\mbox{\sc Cov}_{\mbox{\footnotesize \,The}}(\theta,\theta') = \frac{2}{f_{\mbox{\footnotesize sky}}} \, 
\sum_{\ell \, \ge \, 0} \, \frac{2 \ell + 1}{(4\pi)^2} \, P_{\ell}(\cos \theta) \, P_{\ell}(\cos \theta') \, 
(C_{\ell} + 1/{\bar{n}})^2 \, ,
\end{eqnarray}
where $P_{\ell}$ are the Legendre polynomials, $\bar{n}$ is the number of quasars per steradian, 
and $f_{\mbox{\footnotesize sky}}$ is the fraction of the sky observed in the survey. 
In figure~\ref{fig2} we show the analytical covariance matrices for the linear and non-linear cases. 
For the sake of completeness, in the figure~\ref{fig6} of the Appendix section we compare the error estimates from this theoretical approach, for the linear and non linear cases, to other error estimates, namely, from the jackknife and bootstrap resampling methods.

We calculated the statistical significance of the BAO peak detection with the linear and non-linear 
analytical covariance matrices by following ref.~\cite{Shanks}, that is, we compute the $\chi^2$ 
statistics as a function of the scale dilation parameter $\alpha$, given by 
\begin{equation}\label{chi}
\chi^2(\alpha) = [\omega(\theta) - \omega^{\mbox{\tiny FIT}}(\alpha \theta)]^T \mbox{\sc Cov}^{-1} [\omega(\theta) - \omega^{\mbox{\tiny FIT}}(\alpha \theta)] \, ,
\end{equation}
where $\mbox{\sc Cov}^{-1}$ is the inverse of the covariance matrix; the brackets $[\,]$ and $[\,]^T$ 
represent column and row vectors, respectively; 
$\omega(\theta)$ is the 2PACF data points and $\omega^{\mbox{\tiny FIT}}(\alpha \theta)$ 
is one of two possible best-fit curves to the data: the one given by Eq.~\ref{ajuste}, and the one 
given only by the power law and constant terms, i.e., $C=0$. 
In each case, we fixed the previously best-fit parameters: 
$A, B, C, \gamma, \theta_{\mbox{\tiny FIT}}$, and $\sigma_{\mbox{\tiny FIT}}$, and calculated 
the $\chi^2$ amplitude for each value $\alpha$. 
This was performed in the range $0.25 \le \theta \le 3.5^{\circ}$ and for $0.7 \le \alpha \le 1.2$.

\begin{figure*}[h!]
\centering
\includegraphics[scale=0.65]{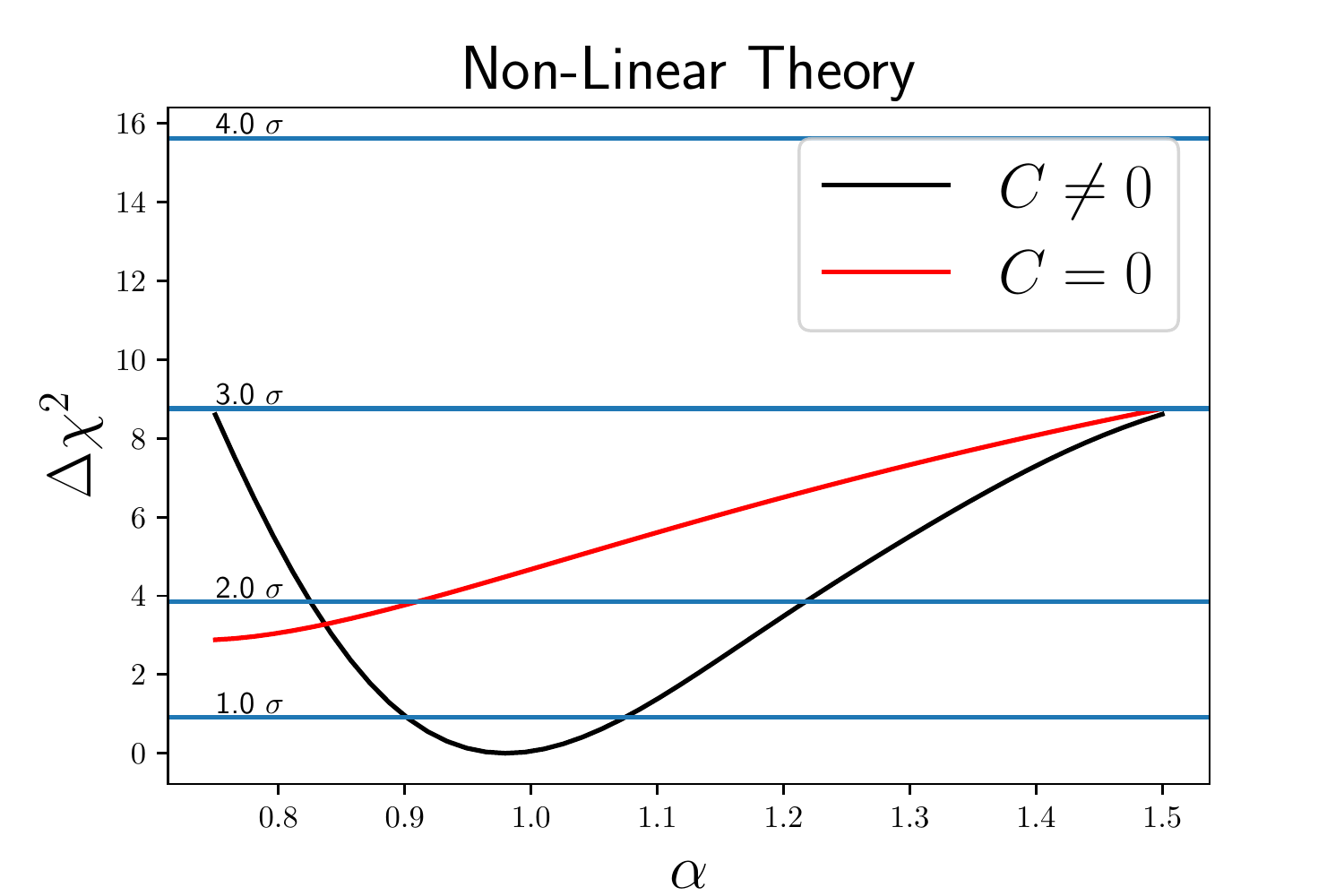}
\caption{
$\chi^2$ analyses: these are the $\Delta\chi^2(\alpha)$ curves obtained calculating $\chi^2(\alpha)$ from Eqs.~\ref{ajuste} and~\ref{chi}, keeping fixed the previously best-fitted parameters, with $C \neq 0$ (with the BAO 
peak; the black curve) and $C = 0$ (without the BAO peak; the red curve).
Here $\Delta\chi^2 = \chi^2(\alpha) \,-\, \chi^2_{min}$, where $\chi^2_{min}$ is the minimum $\chi^2$ value in the case of $C \neq 0$. 
The case presented here corresponds to the analyses using the analytical covariance matrix for the non-linear theory and $N_{\mbox{\scriptsize b}} = 29$, showing a difference of 4.48 among the black and red curves at $\alpha_{min} = 0.9796$, which means that our angular BAO detection has a statistical significance of $2.12 \, \sigma$ (97\% confidence level).
The analyses in the linear theory approach produces the same result (see Table~\ref{table1}).
}
\label{fig3}
\end{figure*}

We show in figure~\ref{fig3} the $\chi^2$ vs. $\alpha$ curves for $C = 0$ (red curve) and $C \neq 0$ (black curve), where $\Delta\chi^2(\alpha) = \chi^2(\alpha) - \chi^2_{min}$, with $\chi^2_{min}$ being the minimum 
$\chi^2$ value for $C \neq 0$. 
The difference between the two $\Delta\chi^2$ curves at $\alpha = \alpha_{min}$, for which the black line attain its minimum value, $\chi^2(\alpha_{min}) = \chi^2_{min}$, provides the statistical significance of the BAO measurement.
As illustrated in figure~\ref{fig3}, for $N_{\mbox{\scriptsize b}} = 29$ and using the non-linear covariance matrix, with a difference of 4.48 between the red and black curves at $\alpha_{min} = 0.9796$, the BAO detection significance is  $2.12 \, \sigma$ (i.e., 97\% confidence level).
A summary of the statistical significances estimated for different $N_{\mbox{\scriptsize b}}$ values and diverse error estimation methods is presented in Table~\ref{table1}; it also includes the jackknife and bootstrap estimators.

\linespread{1.25}
\begin{table}[h!]
\centering
\begin{tabular}{c|cccc}
\hline
Error methods $\backslash$ $N_{\rm b}$  \,\,&\,\, 29 & 33 & 35   \\
\hline
Jackknife         & \,\,2.04$\sigma$ & \,\,1.61$\sigma$ & \,1.95$\sigma$ \\
Bootstrap         & \,\,1.87$\sigma$ & \,\,1.05$\sigma$ & \,1.41$\sigma$ \\ 
Linear Theory     & \,\,2.12$\sigma$ & \,\,1.86$\sigma$ & \,1.81$\sigma$ \\
Non-linear Theory & \,\,2.12$\sigma$ & \,\,1.86$\sigma$ & \,1.81$\sigma$ \\
\hline
\end{tabular}
\linespread{1.0}
\caption{The statistical significance values for different error estimators, for the $N_{b}$ cases 
studied here.
}
\label{table1}
\end{table}

\subsection{Small shifts in the quasars positions} \label{sec:small_shift}

\noindent
To show that the BAO bump we found is a robust detection we have performed an important test, 
namely the \textit{small shifts criterium}, proposed by Carvalho et al. 2016~\cite{Carvalho}. 
This test consists of performing the 2PACF analyses in modified quasars catalogues, that is, we generate a modified catalogue by slightly perturbing the true angular positions of the quasars.  
The main goal of this test is to distinguish the BAO bump, which is expected to be robust and, 
consequently, survive (or be smoothed) under small perturbations in the quasars positions, from those bumps sourced by systematic effects, which should disappear. 

With this in mind, we generated 100 modified catalogues by drawing each quasar's new position from a Gaussian distribution with mean equal to the original position and standard deviation $\sigma_s$, and for each modified catalogue we computed the 2PACF. Our final estimation of the 2PACF for a given scale $\sigma_s$ is the average of all 100 modified catalogues. 
This process was repeated for $\sigma_s=$0.10, 0.20, and 0.30 causing largest displacements of 
$\sim 0.5^{\circ}$, $1.00^{\circ}$, and $1.50^{\circ}$, respectively.  In the calculation of each 2PACF we 
use the same set of 16 random catalogues used in the main analysis, always applying the 
expression~\ref{eq2}. 

\begin{figure}[h]
\centering
\includegraphics[scale=0.65]{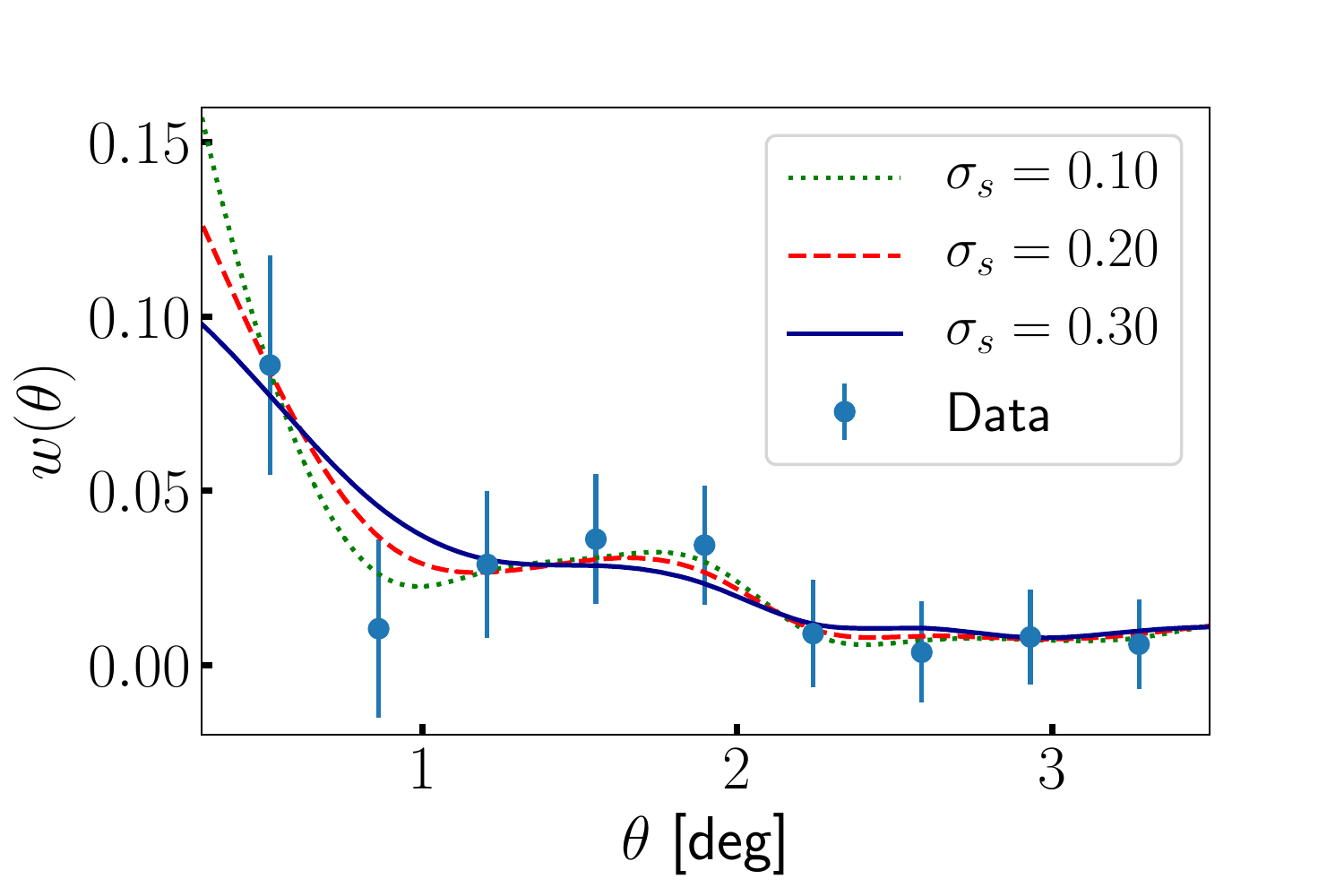}
\vspace{-0.5cm}
\caption{The 2PACF, for $N_{b}=29$, of the perturbed quasar catalogue with angular positions shifted 
by a random amount, following a Gaussian distribution with $\sigma_s = 0.10, 0.20$, and $0.30$ 
(the continuous, dashed and doted lines, respectively). The big dots represent the original data, 
whose error bars were calculated using the analytical covariance matrix estimator (non-linear case).
} 
\label{fig4}
\end{figure}

In figure~\ref{fig4} we display the results achieved in these three cases. 
As observed, the larger the random displacements in the quasars angular positions the smoother the 
2PACF curves, smoothing the BAO bump signature. 
Simultaneously, these displacements also smooth other maxima and minima, possible coming from 
systematic effects or statistical noise, appearing in the original 2PACF. 

\subsection{The projection effect}\label{IVc}

\noindent
To quantify the BAO bump shift due to the projection effect we need to compute the expected 
angular BAO scale, $\theta_{E}^{0}$, which corresponds to the bump position when one computes 
the 2PACF using the equation~(\ref{eq4}) for the case of $\delta z=0$. 
Next, one applies the same procedure but considering $\delta z = 0.05$, which is the thickness of the 
redshift shell used in the actual measurement, to find $\theta_{E}^{\delta z}$. 
Then, the BAO angular scale, $\theta_{\mbox{\footnotesize\sc bao}}$, is given by 
\begin{equation}
\theta_{\mbox{\footnotesize\sc bao}}(\overline{z}) \,=\, \theta_{\mbox{\footnotesize\sc fit}}(\overline{z}) 
+ \Delta \theta(\overline{z}, \delta z) \, \theta_{\mbox{\footnotesize\sc fit}}(\overline{z}) \, , 
\end{equation}
where $\Delta \theta(\overline{z}, \delta z) \equiv (\theta_{E}^{0} - \theta_{E}^{\delta z})/\theta_{E}^{0}$, 
that is, computing the expected values $\theta_{E}^{0}$ and $\theta_{E}^{\delta z}$, we 
perform the shift in the measured value $\theta_{\mbox{\footnotesize\sc fit}}(\overline{z})$ to find 
the BAO angular scale $\theta_{\mbox{\footnotesize\sc bao}}(\overline{z})$. 

This numerical analysis assumes six cosmological parameters: the baryon density 
$\omega_b \equiv \Omega_b h^2$, the cold dark matter density $\omega_c \equiv \Omega_c h^2$, 
the ratio between the sound horizon and the angular diameter distance to decoupling $\Theta$, the 
optical depth to reionization $\tau$, the overall normalization of the primordial power spectrum 
$A_s$, and the tilted scalar spectral index $n_s$. 
As a reference model we use 
$\omega_b = 0.0226,\, \omega_c = 0.112,\, 100\, \Theta=1.04,\, \tau=0.09,\, 10^{9}\,A_s = 2.2,\, 
n_s = 0.96$, and $h$ is the Hubble constant in unit of $ 100\, km\,s^{-1}\,Mpc^{-1}$. 
We also assume flatness and, for neutrinos, the effective number of relativistic degrees of freedom 
is equal to $3.046$.

Using this reference model we find the values $\theta_{E}^0=1.518^\circ$ for $\delta z = 0$
and $\theta_{E}^{\delta z}=1.504^\circ$ for $\delta z = 0.05$. 
The shift obtained is $0.014^\circ$, i.e., the projection effect is almost negligible since it produces 
a shift in the BAO angular position of $\Delta \theta(\overline{z}, \delta z)\,=\,0.9\%$.

\subsection{Cosmological constraints}\label{CC}

\noindent
Here we measured a new angular acoustic scale, $\theta_{\mbox{\footnotesize\sc bao}}$ at 
$\overline{z} = 2.225$. 
We combine this new measurement with other eight data points provided by~\cite{Carvalho,Alcaniz} 
(obtained following the same quasi-independent model approach) to constrain parameters in the 
cosmological models with constant and variable equations of state, namely wCDM and w$(a)$CDM 
models. 

\noindent
For the w$(a)$CDM model we assume the Barboza-Alcaniz parametrized equation of 
state~\cite{BarbozaAlcaniz}: 
${\rm w}(a) = {\rm w}_0 + {\rm w}_a [(1- a) / (2a^2 - 2a + 1)]$. 
The relation between the angular BAO scale, $\theta_{\mbox{\footnotesize\sc bao}}$, and the angular 
diameter distance, $D_A$, is obtained using the expression 
\begin{equation}
 \theta_{\rm BAO}(z) = \frac{r_s(z_{drag})}{(1+z) \, D_A(z)} \, ,
\end{equation} 
where $r_s(z_{drag})$ is the comoving sound horizon at the end of baryon drag, and is provided by CMB 
measurements. 
To constrain the cosmological parameters we combine the transversal BAO data with the CMB shift 
parameter information defined as $R \equiv \sqrt{\Omega_m H_0^2} \,\, r_s(z_{rec})$, where 
$r_s(z_{rec})$ is the comoving sound horizon at recombination. 
In our case, we use the shift parameter given by the Planck collaboration 
$R = 1.7407 \pm 0.0094$~\cite{Wang2013}. 

\begin{figure}[h!]
\hspace{-0.2cm}
\includegraphics[width=0.56\textwidth]{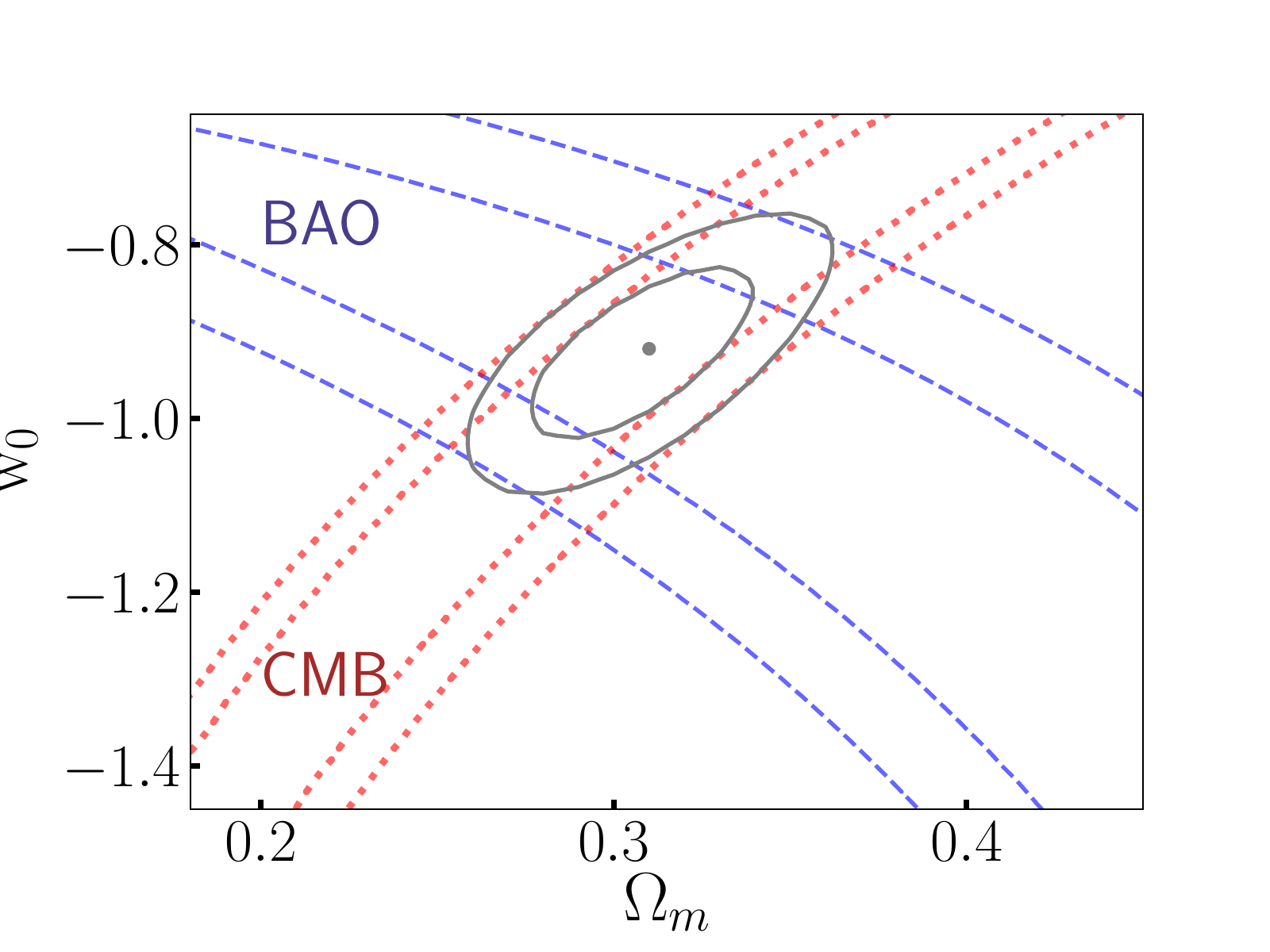}
\hspace{-1cm}
\includegraphics[width=0.56\textwidth]{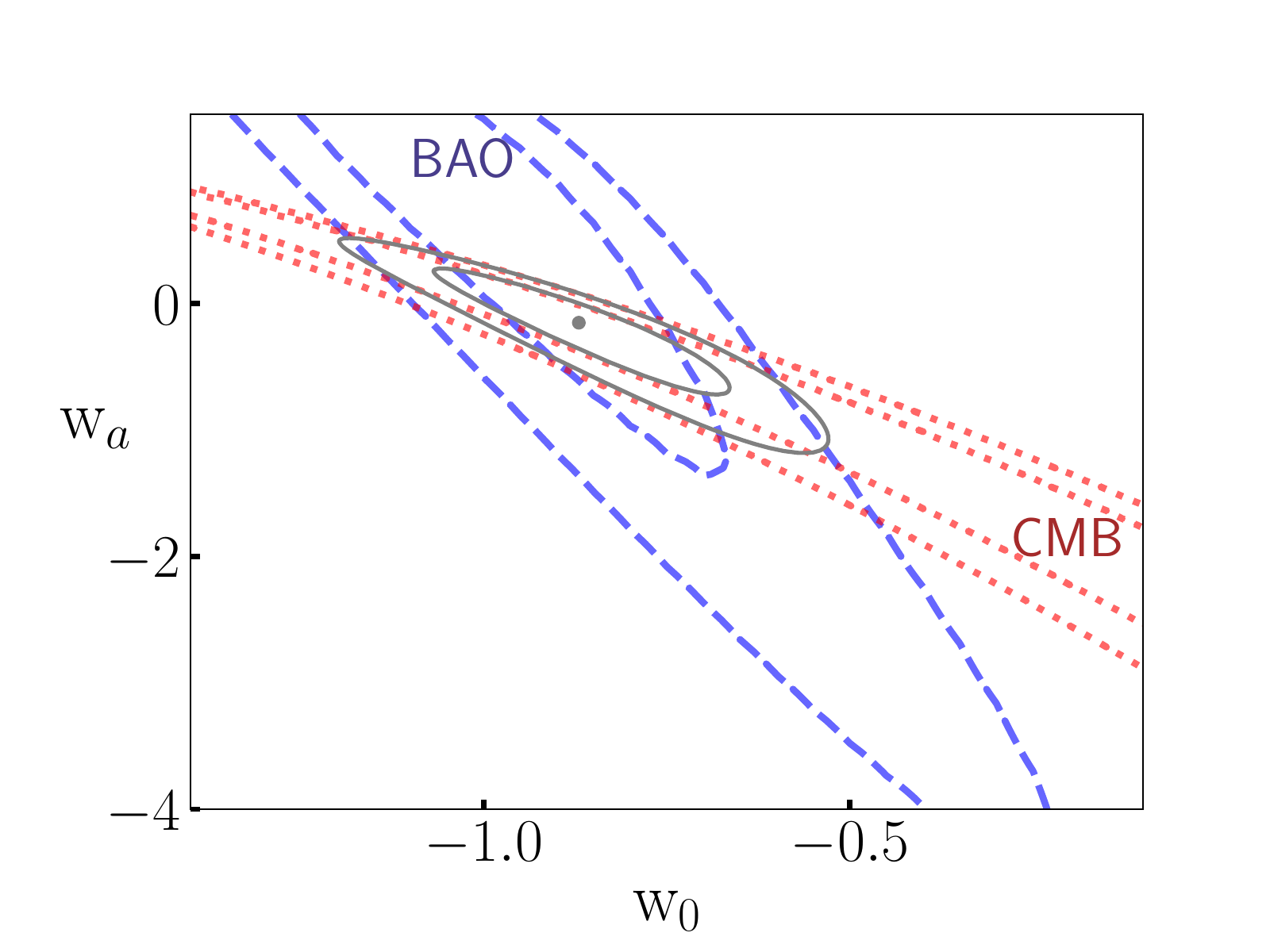}
\caption{One and two $\sigma$ confidence level contours of the $\Omega_m-{\mbox w}_0$ and ${\mbox w}_0-{\mbox w}_a$ 
planes for the wCDM and w$(a)$CDM models, respectively.} 
\label{contours}
\end{figure}

Assuming the sound horizon value obtained by WMAP9~\cite{wmap9}, $r_s(z_{drag}) = 106.61 \pm 3.47$ Mpc/h, 
jointly to the Planck measurement of the shift parameter\footnote{We use these values from different 
surveys because they are expected to be non-correlated data, therefore the current analysis is valid.}, 
we constrain the $\Omega_m$ and w$_0$ for the wCDM model. 
Thus, the best-fit corresponds to $\Omega_m= 0.31 \pm 0.02$ and w$_0 = -0.92 \pm 0.06$ 
for the wCDM model, while for w$(a)$CDM we found w$_0= -0.87\pm 0.13$ and 
w$_a= -0.15\pm 0.32$. 
The confidence level contours for $\Omega_m-\mbox{\rm w}_0$ for wCDM model and w$_0-$w$_a$ 
for w$(a)$CDM model can be seen in figure~\ref{contours}.

\section{\label{sec5} Conclusions and Final Remarks} 

Following the same quasi model-independent approach used in the 
refs.~\cite{Carvalho,Alcaniz,Gabriela2,Carnero,Sanchez11}, we obtained a robust and moderately 
significant measurement of the BAO angular scale, 
$\theta_{\mbox{\footnotesize\sc bao}} = 1.77^{\circ} \pm 0.31^{\circ}$ at $\overline{z}=2.225$, 
using the excellent catalogue DR12Q of the SDSS collaboration. 
The statistical significance of this measurement, $2.12 \,\sigma$, i.e., 97\% confidence level, was 
calculated with the non-linear theoretical covariance matrix expected in the $\Lambda$CDM 
concordance model (see section~\ref{errors}), a result that is in excellent agreement with reported 
measurements (see, e.g., refs.~\cite{Carnero,Crocce11b}).

After measuring the acoustic peak position through the 2PACF, it was necessary to apply a shift 
correction due to unavoidable projection effects caused by the finite thickness of the redshift shell, 
$\delta z \ne 0$, of the quasars sample. 
But, as expected from previous analyses~\cite{Sanchez11}, the numerical evaluation of this effect 
showed a negligible shift correction of less than 1\%, evidencing a weak dependence of our 
measurement with respect to the reference cosmological model. 
At the same time, one observes that the effective model contribution in this approach comes 
from the sound horizon value, $r_s$, which is derived from CMB measurements in a 
model-dependent way~\cite{wmap9}.

Additionally, we validated our angular BAO detection through diverse robustness tests. 
In fact, we show in sections~\ref{bin-size} and~\ref{errors} that the BAO signal is also robust under 
different bin-size choices, $N_{b}$, used in the calculation of the 2PACF and the corresponding BAO 
scale measurement. 
Additionally, we verify in section~\ref{sec:small_shift} that our measurement is stable under small 
Gaussian random perturbations of the angular positions of the quasars sample (this is a useful 
procedure which also leads to discriminate possible noise bumps from the true BAO signature).

With this angular BAO measurement we have increased the number of available BAO angular 
scale data, this time with a measurement at high redshift. 
We combined this new measurement with other eight data points (from 
refs.~\cite{Carvalho,Alcaniz}) to constrain the cosmological parameters of wCDM and w$(a)$CDM 
models. 
Our results show an excellent agreement with the standard cosmological model 
$\Lambda$CDM: the best-fit corresponds to 
$\Omega_m= 0.31 \pm 0.02$ and w$_0 = -0.92 \pm 0.06$ for the wCDM model, 
while for w$(a)$CDM we found w$_0= -0.87 \pm 0.13$ and w$_a= -0.15 \pm 0.32$ (see 
figure~\ref{contours} and section~\ref{CC} for details).

Moreover, it is worth emphasizing that our likelihood analyses considers the covariance matrices 
sourced by linear and non-linear analytical power spectra, finding a negligible difference between 
both cases, an expected result since at high redshift quasar clustering would be little affected 
by non-linear physics. 
Finally, for the sake of completeness, the likelihood analysis was extended to other well-known 
approaches, such as jackknife and bootstrap methods. 
In this way we compared the outcomes of diverse error estimators, confirming that jackknife and 
bootstrap resampling methods overestimate the 2PACF error bars while pure random catalogues 
approach underestimate them. These analyses are presented in the Appendix section.

\section*{Acknowledgements}
We are grateful to Joel C. Carvalho for his support in numerical programming, and to him and Jailson 
Alcaniz for insightful discussions. 
The authors thank the CAPES PVE project 88881.064966/2014-01, within the {\em Science 
without Borders Program}. 
CPN, GCC, and HSX also acknowledge the Brazilian funding agencies FAPERJ (P\'os-doc Nota 10), 
CNPq, and FAPESP, respectively, for the financial support.

\appendix
\section{The jackknife and bootstrap error estimators}

\noindent
We estimate the error bars of the 2PACF using the well-known jackknife and bootstrap resampling 
methods~\cite{Myers,Ross,Shanks,Norberg}. 

For the jackknife procedure, we first create 35 quasars jackknifed samples using the original 
DR12Q studied here, and then we measure the 2PACF in each one of them. 
The error bars are calculated using the covariance matrix, which is constructed according to eq. (6) 
of ref.~\cite{Crocce11b}. 

For the bootstrap method, we consider $N_{sub} \,=\, 47$ sub-samples and we resample as in~\cite{Norberg}: 
$N_{r} \,=\, 3 N_{sub}$, ($N_{r}$ is the number of random sub-samples with replacement from the original set) then we generate an amount of 100 bootstrap samples. 
The error bars are calculated using the covariance matrix, constructed following 
Norberg et al. (2009)~\cite{Norberg}. 
 
\begin{figure}[h]
\centering
\includegraphics[width=10.8cm, height=7.7cm]{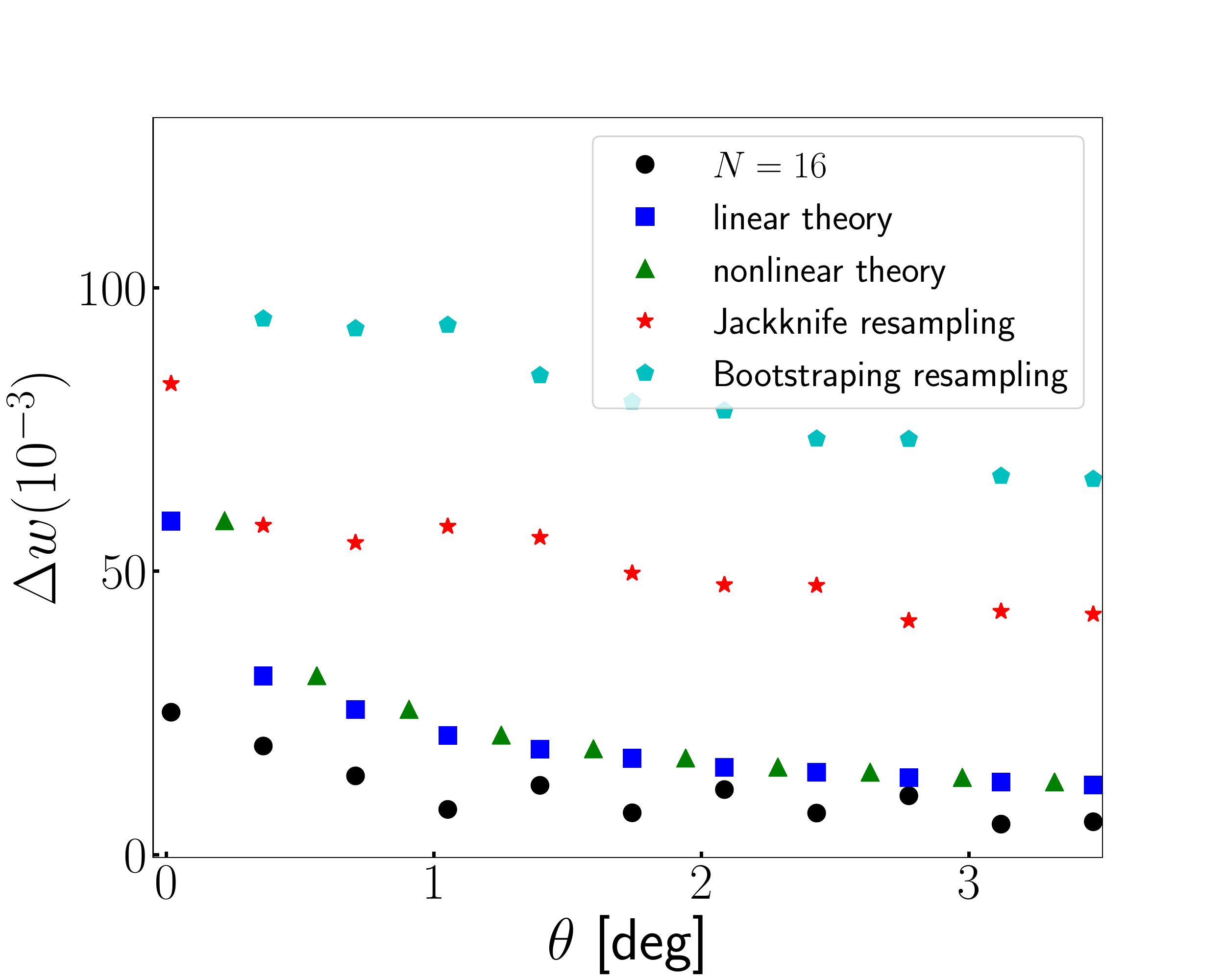}
\caption{
Comparison of the diverse error estimation methods. 
The triangles representing the non-linear theory approach were slightly shifted to the right side.
} 
\label{fig6}
\end{figure}

As already reported in the literature~\cite{Crocce11a,Crocce11b,Norberg}, the error bars for the 
2PACF are over-estimated by the jackknife and bootstrap error estimators, as compared with the 
analytical covariance matrix error calculation. 
In figure~\ref{fig6} we compare the error bars obtained from diverse estimators, where one 
can observe that the jackknife and bootstrap methods actually over-estimate the errors of the 
2PACF. 
The black dots represent the error bars estimated directly from the $N=16$ random catalogues, 
that is, as the standard deviation from the set of 2PACF obtained with each one of these random 
catalogues.

\section{The Random catalogues}

\noindent
A crucial part in the robust, and statistically significant, detection of the BAO angular scale is the 
production of a set of random catalogues with the same features as the observed quasar catalogue, apart from clustering ones: 
they should contain cosmic objects homogeneously distributed in the same geometry of the survey, and their number 
should be equal to the number of objects in the data catalogue. 

\begin{figure}[h]
\centering
\mbox{\hspace{-0.2cm}
\includegraphics[scale=0.7]{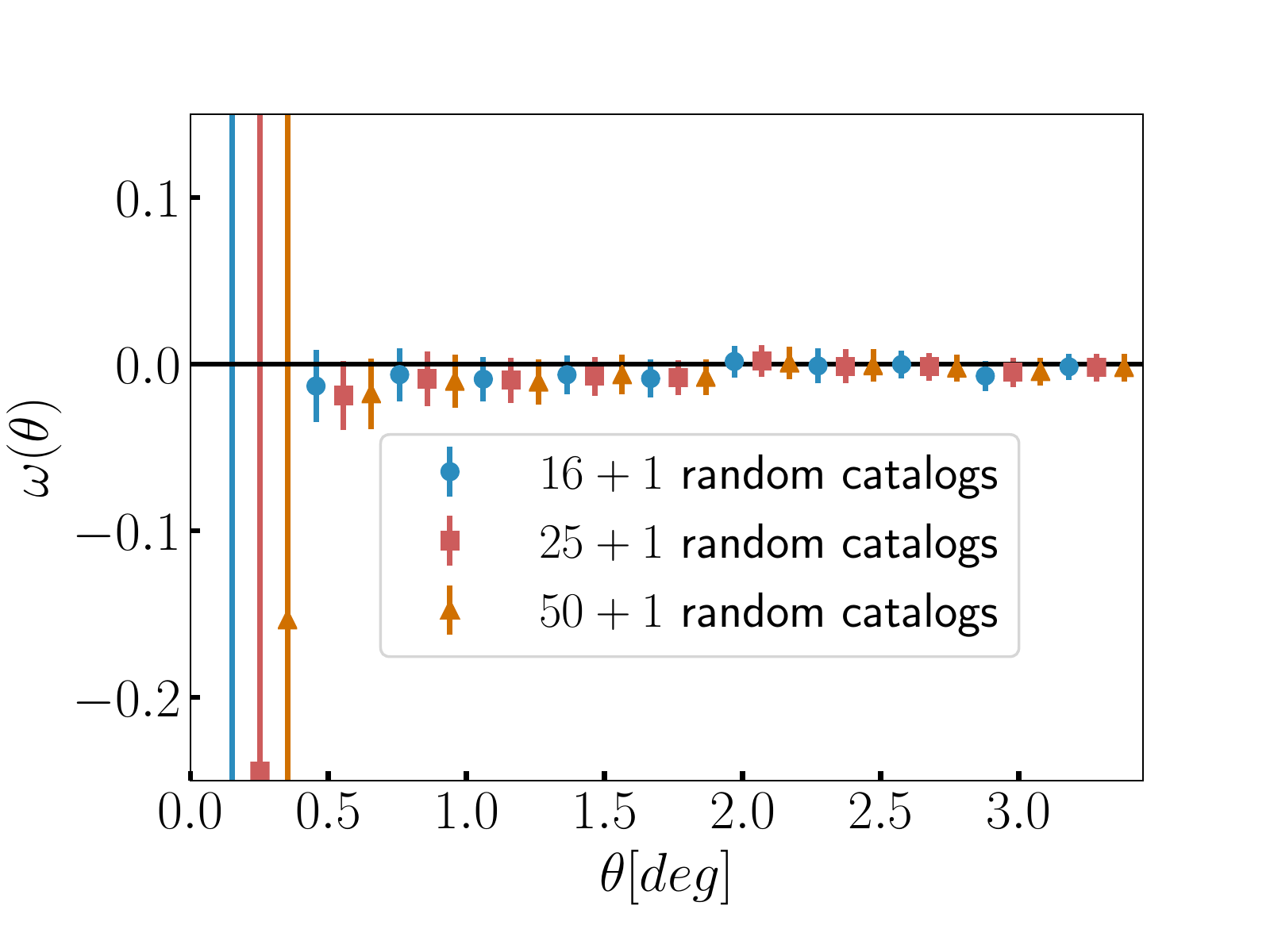}
}
\caption{The 2PACF ($N_b = 29$) obtained using $N=$ 16 (dots), 25 (squares), and 50 (triangles) random catalogues, 
where the last random catalogue in each sample was considered as the {\it pseudo-data-catalogue} 
(see the text for details). 
The square and triangle markers were artificially shifted to the right by $0.10^{\circ}$ and $0.20^{\circ}$, 
respectively, for clarity. 
Here, the error bars were obtained using the random catalogues (i.e., they are the standard 
deviations of the $N$ 2PACF calculated), which according to the results of the Appendix A, illustrated 
in figure~\ref{fig6}, are smaller than the correct ones (obtained from the analytical covariance matrix).}
\label{fig7}
\end{figure}

The methodology used to generate our random catalogues in order to obtain a homogeneous Poisson 
sampling of cosmic objects~\cite{Peebles-Hauser} was to shuffle the data angular coordinates in the shell in a way that any 
possible correlation will be destroyed~\cite{Rego}. 
There are many approaches to produce these simulated data, but perhaps the most important task is 
to certify that they are indeed featureless, and the simplest way to do this is through a {\it null test} (see, e.g., 
section 5 in ref.~\cite{Landy-Szalay}). 
Consider that we have produced $N+1$ random catalogues. 
A null test assumes any of the random catalogues as the {\it pseudo-data catalogue} and estimates the 
2PACF, using equation (\ref{2PACF}), with the remaining $N$ random catalogues. 
Since there is no preferred clustering at any scale, the expected 2PACF is, up to the error bars, 
a null correlation, that is, one does not expect angular correlations at any scale for the 
{\it pseudo-data catalogue}. 
We performed this test considering three cases: $N=$ 16, 25, and 50, and our results are displayed in 
figure~\ref{fig7}, where we observe that they show an excellent performance.

\end{document}